\documentstyle[aps, preprint]{revtex}
\tightenlines
\draft
\begin{document}
\title{Extra force and extra mass
from noncompact Kaluza-Klein theory in a cosmological model}
\author{$^1$Jos\'e Edgar Madriz Aguilar\footnote{
E-mail address: edgar@itzel.ifm.umich.mx}
and $^{2}$Mauricio Bellini\footnote{
E-mail address: mbellini@mdp.edu.ar}}
\address{$^1$Instituto de F\'{\i}sica y Matem\'aticas,
AP: 2-82, (58040) Universidad Michoacana de San Nicol\'as de Hidalgo,
Morelia, Michoac\'an, M\'exico.\\
$^2$Departamento de F\'{\i}sica, Facultad de Ciencias Exactas y Naturales,
Universidad Nacional de Mar del Plata and
Consejo Nacional de Ciencia y Tecnolog\'{\i}a (CONICET),
Funes 3350, (7600) Mar del Plata, Argentina.}

\vskip .2cm
\maketitle
\begin{abstract}
Using the Hamilton-Jacobi formalism,
we study extra force and extra mass in a recently introduced
noncompact Kaluza-Klein cosmological model. We examine the inertial
4D mass $m_0$ of the inflaton field
on a 4D FRW bulk in two examples.
We find that $m_0$ has a geometrical origin and
antigravitational effects on a non inertial 4D bulk
should be a consequence of the motion of the fifth coordinate with
respect to the 4D bulk.
\end{abstract}
\vskip .2cm
\noindent
Pacs numbers: 04.20.Jb, 11.10.kk, 98.80.Cq \\
\vskip .1cm
\section{Introduction}

The extra dimension is already known to be of potential importance for
cosmology\cite{uno,dos}. During the last years there were many attempts
to construct a consistent brane world (BW)
cosmology\cite{.}. On the other hand,
the induced-matter,
or space-time-matter (STM) theory
stands out for its closeness to the Einstein's project of considering
matter and radiation as manifestations of pure geometry\cite{eins}.
Indeed,
the gist of the whole theory is to assert that, by embedding the ordinary
space-time into a five-dimensional vacuum space, it is possible to describe
the macroscopic properties of matter in geometrical terms. Picking up several
examples of cosmological and gravitational models, the theory shows how
interpret the energy-momentum tensor corresponding to some standard matter
configurations in terms of the geometry of the five-dimensional vacuum-space.
It was recently questioned in a recent article\cite{anderson}.
The induced matter theory is sometimes given the name of Kaluza-Klein
noncompact field gravity, since Klein's compactness condition\cite{2}
is dropped from the basic assumptions of the theory.

The idea that matter in four dimensions (4D) can be explained from
a 5D Ricci-flat ($R_{AB}=0$) Riemannian manifold is a consequence of the
Campbell's theorem. It says that any analytic $N$-dimensional Riemannian
manifold can be locally embedded in a $(N+1)$-dimensional Ricci-flat
manifold. This is of great importance for establishing the generality of the
proposal that 4D field equations with sources can be locally embedded
in 5D field equations without sources\cite{10}.
On the other hand, recently was remarked by Anderson that such that
theorem lends only inadequate support, both because it offers no guarantee
of continuous dependence on the data and because it disregards
causality\cite{anderson}. In his paper he point out that the theorem is only
valid for analytic functions which renders it inappropiate for the study of the
relativistic field equations in modern physics. However, analytic
functions seems only inappropiate to describe some topological defects
like black holes, but not to study cosmological models for which the
manifold is global and there are no singularities.
Other recent contribution to this issue was made by Katzourabis\cite{kat},
who demonstrated a global generalization of the Campbell's theorem, removing
the asumption of locality:
{\em Any pseudo-Riemannian analytic $N$-manifold can be embedded (the
whole structure) naturally and isometrically into $(N+1)$-bulk with
arbitrarily prefixed topological structure, fibred over the given, with
vanishing Ricci curvature and torsion of some connection locally compatible
with a global metric on the bulk}.
More generally he found that {\em any 4D-manifold with a global tensor field
representing the solution of 4D General Relativity equations can be
embedded naturally and isometrically into (4+d)-manifold ($d \in\mathbf{N}$)
Ricci flat and torsion free, where the global topology, the differential
structure and the dimensionality of the bulk can be fixed arbitrarily.}
In few words, 4D phenomenological
matter can be induced by a 5D apparent vacuum in the framework of
cosmological models where the universe is described by
a 3D spatially isotropic, homogeneous and flat (analytic) metric.

An important question in STM and BW theories can be split up in two parts.
1) How to extract the correct 4D interpretation from geometrical objects,
like scalar fields which appear in more than four dimensions. 2) How to
predict observable effects from the extra dimensions.
There are at least two ways to measure the effect, one which
shows no change\cite{15} and one which does\cite{otra}.
For example, an attempt
to understand  the first of these questions in the framework of
inflationary cosmology
[which is governed by the neutral scalar (inflaton) field],
from a 5D flat Riemannian manifold
was made in\cite{NPB}.

In the last years extra force and extra mass has been subject of study
\cite{14}.
It should be an observable effect from extra dimensions on the
4D spacetime.
The aim of this work is to extend the Hamilton-Jacobi formalism
developed by Ponce de Leon\cite{15} to cosmological models where
the expansion of the universe is governed by a single inflaton field.
This interpretation has the advantage of being free of the complications
and ambiguities of the geodesic approach.
The formalism provides an unambiguous expression for the rest mass and
its variation along the motion as observed in 4D. It is
independent of the coordinates and any parametrization used along the motion.

\section{Cosmological approach}

In a cosmological context, the energy of scalar fields has been argued
to contribute to the expansion of the
universe\cite{jd}, and has been proposed to
explain inflation\cite{Guth} as well as the presently accelerated
expansion\cite{we}.
In this paper we shall consider a cosmological model governed by a neutral
scalar field that initially suffers an inflationary expansion that has a change
of phase towards a decelerated (radiation and later matter dominated)
expansion that thereinafter evolves towards the observed present day
(quintessential) expansion.

We consider the recently introduced\cite{PLB}
5D metric
\begin{equation}\label{6}
dS^2 = \epsilon\left(\psi^2 dN^2 - \psi^2 e^{2N} dr^2 - d\psi^2\right),
\end{equation}
where $dr^2 = dx^2+dy^2+dz^2$. In order to describe cosmological models
we shall consider a 3D spatially isotropic, homogeneous and flat metric
(i.e., $x^2=^2=z^2$).
Here, the coordinates ($N$,$\vec r$)
are dimensionless, the fifth coordinate
$\psi $ has spatial unities and $\epsilon$ is a dimensionless parameter
that can take the values $\epsilon = 1,-1$.
The metric (\ref{6}) describes a
flat 5D manifold in apparent vacuum ($G_{AB}=0$).
With the aim to describe neutral matter in a 5D geometrical vacuum
(\ref{6}), as in the paper\cite{PLB1}, we can consider the Lagrangian
\begin{equation}\label{1}
^{(5)}{\rm L}(\varphi,\varphi_{,A}) =
-\sqrt{\left|\frac{^{(5)}
g}{^{(5)}g_0}\right|} \  ^{(5)}{\cal L}(\varphi,\varphi_{,A}),
\end{equation}
where $|^{(5)}g|=\psi^8 e^{6N}$
is the absolute value of the determinant for the 5D metric tensor with
components $g_{AB}$ ($A,B$ take the values $0,1,2,3,4$) and
$|^{(5)}g_0|=\psi^8_0 e^{6N_0}$
is a constant of dimensionalization determined
by $|^{(5)}g|$ evaluated at $\psi=\psi_0$ and $N=N_0$.
We shall consider $N_0=0$, so that
$^{(5)}g_0=\psi^8_0$.
We define the vacuum
as a purely kinetic 5D-lagrangian on a globally 5D-flat metric [in our
case, the metric (\ref{6})]\cite{PLB1}.
Since the 5D metric (\ref{6}) describes a manifold in apparent
vacuum, the density Lagrangian
${\cal L}$ in (\ref{1}) must to be
\begin{equation}\label{1'}
^{(5)}{\cal L}(\varphi,\varphi_{,A}) =
\frac{1}{2} g^{AB} \varphi_{,A} \varphi_{,B},
\end{equation}
which represents a free scalar field.
In the 3D comoving frame $U^r=0$,
the geodesic dynamics ${dU^C \over dS}=-\Gamma^C_{AB} U^A U^B$
with $g_{AB} U^A U^B=1$, give us the velocities
\begin{equation}\label{ui}
U^{\psi} = - {1 \over \sqrt{u^2(N)-1}}, \qquad U^{r}=0, \qquad
U^N={u(N) \over \psi\sqrt{u^2(N)-1}},
\end{equation}
which are satisfied for $S(N)=\pm |N|$.
We shall consider the case $S(N) = |N|$.
In this representation ${d\psi \over dN}=\psi/u(N)$.
Thus the fifth coordinate evolves as
\begin{equation}\label{psi}
\psi(N) = \psi_0 e^{\int dN/u(N)}.
\end{equation}
Here, $\psi_0$ is a constant of integration that has spatial unities.
From the mathematical point of view, we are taking a foliation
of the 5D metric (\ref{6}) with $r$ constant.

In order to obtain equations of practical use, we can introduce the
action ${\cal S}(x^A)$ as a function of the generalyzed coordinates
$x^A$. Hence, since the momentum
$P^A = - {\partial {\cal S} \over \partial x^A}$,
for a diagonal tensor metric $g^{AB}$ we obtain the Hamilton-Jacobi
equation
\begin{equation}\label{HJ}
g^{AB}  \left(\frac{\partial {\cal S}}{\partial x^A}\right)
\left(\frac{\partial {\cal S}}{\partial x^B}\right)= M^2_{(5)},
\end{equation}
where $M_{(5)}$ is the invariant 5D gravitational
mass of the object under study (in our case, the mass of the inflaton
field).
In the particular frame (\ref{ui}), with the Lagrangian
(\ref{1}) and (\ref{1'}), $M_{(5)}$ describes the 5D mass of the
scalar field $\varphi$ in the frame (\ref{ui}).
In this case the tensor metric is symmetric (and diagonal), and
the Hamilton-Jacobi equation (\ref{HJ}) adopts the particular
form
\begin{equation}
g^{NN} \left(\frac{\partial {\cal S}}{\partial \varphi_{,N}}\right)^2 +
g^{\psi\psi} \left(\frac{\partial {\cal S}}{\partial
\varphi_{,\psi}}\right)^2 =
M^2_{(5)}.
\end{equation}

\section{Extra force and extra mass}

In general, the line element (\ref{6}) can be
written as:
\begin{equation}
dS^2 = ds^2 + dS^2_{(4)},
\end{equation}
where $ds^2$ describes the 4D line element and $dS^2_{(4)}$ only
the line element related with the fifth coordinate.
We shall define the extra force
\begin{equation}
F^{ext} = \frac{dP^{x^4}}{ds},
\end{equation}
as the force on the sub manifold
$ds^2$ due to the motion of the fifth coordinate.
In general, $P^{x^4}$ is defined as
\begin{equation}
P^{x^4} = \frac{\partial ^{(5)}L}{\partial \varphi_{,x^{4}}}.
\end{equation}
In the frame (\ref{ui}) $P^{x^4} \equiv P^{\psi}$, and
is given by
$P^{\psi} = - {\psi^4 e^{3N}\over \psi^2_0} \left(g^{\psi\psi}\right)^2
\varphi_{,\psi}$, which also can be written in terms of the potential
\begin{equation}
P^{\psi} = -\frac{\psi^4 e^{3N}}{\psi^2_0}
g^{\psi\psi} \frac{\partial V(\varphi)}{\partial\varphi_{,\psi}},
\end{equation}
where $V(\varphi)$ depends on the particular frame of the observer. In the
next subsections we shall see different forms of it in two different
frames.
Hence, in the frame (\ref{ui}) the extra force holds
\begin{equation}
F^{ext} = \frac{\psi^3 e^{3N}}{\psi^2_0}
\left( 3 \frac{\stackrel{\star}{\psi}}{\psi} \varphi_{,\psi}
+ 3 \varphi_{,\psi}+ \stackrel{\star}{\varphi}_{,\psi}\right),
\end{equation}
where the overstar denotes the derivative with respect to $N$.

On the other hand, from the
equation $g_{AB} U^A U^B =1$, we obtain the
invariant 5D mass $M_{(5)}$
\begin{equation}
g_{AB} P^A P^B =M^2_{(5)},
\end{equation}
where $P^A= M_{(5)} U^A$ is the momentum.
For example, in the frame (\ref{ui}) the 4D mass $m_0$ and the
5D invariant mass $M_{(5)}$ are given respectively by
\begin{equation}
M^2_{(5)}= g^{NN} \left(\frac{\partial {\cal S}}{\partial \varphi_{,N}}\right)^2+
g^{\psi\psi} \left(\frac{\partial {\cal S}}{\partial \varphi_{,\psi}}\right)^2 ,
\qquad m^2_0 = g^{NN} \left(\frac{\partial {\cal S}}{\partial \varphi_{,N}}\right)^2,
\end{equation}
so that its diference
\begin{equation}
m^2_0 - M^2_{(5)} = -g^{\psi\psi}
\left(\frac{\partial {\cal S}}{\partial \varphi_{,\psi}}\right)^2,
\end{equation}
is nonzero. The interesting here is that $m^2_0 > M^2_{(5)}$. In other
words, in the frame (\ref{ui})
the motion of the fifth coordinate has an antigravitational effect
on the field $\varphi$ on the submanifold (or bulk) $ds^2$.
In the next sections we shall study some
examples which could be relevant in cosmological models.

\subsection{A 4D FRW cosmology}

We consider the transformations:
$t = \int \psi(N) dN$, $R=r\psi$, $ L= \psi(N) \  e^{-\int dN/u(N)}$,
such that for $\psi(t)=1/h(t)$,
we obtain the 5D metric
$dS^2 = \epsilon\left(dt^2 - e^{2\int h(t) dt} dR^2 - dL^2\right)$,
where $L=\psi_0$ is a constant and $h^2(t)=\left(\dot b/b\right)^2 = (8/3)\pi G
\left<\rho\right>$
is the effective
Hubble parameter defined from the effective scale factor of the
universe $b$.
In this frame $(R,t,L)$, the
velocities $ \hat U^A ={\partial \hat x^A \over \partial x^B} U^B$,
are
\begin{equation} \label{10}
U^t=\frac{2u(t)}{\sqrt{u^2(t)-1}}, \qquad
U^R=-\frac{2r}{\sqrt{u^2(t)-1}}, \qquad U^L=0,
\end{equation}
where
the old velocities $U^B$ are $U^N$, $U^r=0$ and $U^{\psi}$
and the velocities $\hat U^B$ are constrained by the condition
\begin{equation}\label{con}
\hat g_{AB} \hat U^A \hat U^B =1.
\end{equation}
The variables $(t,R,L)$ has physical meaning, because
$t$ is the cosmic time and $(R,L)$ are spatial variables.
Since the line element is a function of time $t$
(i.e., $S\equiv S(t)$), the new coordinate $R$ give us
the physical distance between galaxies separated
by cosmological distances: $R(t)=r/h(t)$.
Note that for $r >1$ ($r <1$), the 3D spatial distance $R(t)$ is defined
on super (sub) Hubble scales.
Furthermore $b(t)$ is the effective scale factor of the
universe and describes its effective 3D Eucliden (spatial) volume.
Hence, the effective 4D metric
is a spatially (3D) flat FRW one
\begin{equation}\label{frw}
dS^2 \rightarrow ds^2 = \epsilon
\left(dt^2 - e^{2\int h(t) dt} dR^2\right),
\end{equation}
and has a effective
4D scalar curvature $^{(4)}{\cal R} = 6(\dot h + 2 h^2)$. The
metric (\ref{frw}) has a metric tensor with components $g_{\mu\nu}$
($\mu,\nu$ take the values $0,1,2,3$).
The absolute value of the determinant for this tensor is $|^{(4)}g|
=(b/b_0)^6$.
The density Lagrangian in this new frame was obtained in a previous
work\cite{PLB1}
\begin{equation} \label{aa}
^{(4)} {\cal L}\left[\varphi(\vec{R},t), \varphi_{,\mu}(\vec{R},t)\right]
= \frac{1}{2} g^{\mu\nu} \varphi_{,\mu} \varphi_{,\nu}
- \frac{1}{2} \left[(R h)^2  -
\frac{b^2_0}{b^2} \right] \  \left(\nabla_R \varphi\right)^2,
\end{equation}
and the equation of motion for $\varphi$ yields
\begin{equation}\label{bb}
\ddot\varphi  +  3 h\dot\varphi -\frac{b^2_0}{b^2} \nabla^2_R \varphi
+ \left[\left(4\frac{h^3}{\dot h} - 3\frac{\dot h}{h}
- 3\frac{h^5}{\dot h^2}\right) \dot\varphi +
\left( \frac{b^2_0}{b^2} - h^2 R^2\right)\nabla^2_R\varphi\right]=0.
\end{equation}
From eqs. (\ref{aa}) and (\ref{bb}), we obtain respectively
the effective scalar 4D potential
$V(\varphi)$ and its derivative with respect
to $\varphi(\vec{R},t)$
\begin{eqnarray}
V(\varphi) & \equiv & \frac{1}{2}\left[ (R h)^2
- \left(\frac{b_0}{b}\right)^2 \right] \left(\nabla_R\varphi\right)^2,
\label{au} \\
V'(\varphi) & \equiv  &
 \left(4\frac{h^3}{\dot h} - 3\frac{\dot h}{h} -
3\frac{h^5}{\dot h^2}\right) \dot\varphi
+\left(\frac{b^2_0}{b^2} - h^2 R^2\right)\nabla^2_R\varphi,\label{a1}
\end{eqnarray}
where the prime denotes the derivative with respecto to $\varphi $.
The equations (\ref{aa}) and (\ref{bb}) describe the dynamics of the inflaton field
$\varphi(\vec{R},t)$ in a metric (\ref{frw}) with a Lagrangian
\begin{equation}\label{l4}
^{(4)}{\cal L}[\varphi(\vec{R},t),\varphi_{,A}(\vec{R},t)] =
-\sqrt{\left|\frac{^{(4)}g}{^{(4)}g_0}\right|}
\left[\frac{1}{2} g^{\mu\nu} \varphi_{,\mu}\varphi_{,\nu}
+V(\varphi)\right],
\end{equation}
where $\left|^{(4)}g_0\right|=1$.

Furthermore, the function $u$ can be written as a function of time
$u(t) = -{h^2 \over \dot h}$,
where the overdot represents the derivative with respect to the time.
The solution $N={\rm arctanh}[1/u(t)]$ corresponds to a
time dependent power-law expanding universe
$h(t)=p(t) t^{-1}$, such that the effective scale factor go as
$b \sim e^{\int p(t)/t dt}$. A model for the evolution of the universe
was recently developed in \cite{ultimo}.
Furthermore, the 4D energy
density $\rho$ and the pressure ${\rm p}$ are\cite{PLB}
\begin{eqnarray}
&& 8 \pi G \rho = 3 h^2,\\
&& 8\pi G {\rm p} = -(3h^2 + 2 \dot h).
\end{eqnarray}
Note that
the function $u(t)$ can be related
to the deceleration parameter $q(t)=-\ddot b b/\dot b^2$: $u(t)=1/[1+q(t)]$.
In what follows we shall consider $\epsilon =1$ when the universe
is accelerated ($q < 0$) and $\epsilon = -1$ in epochs when it is decelerated
($q>0$).
From the condition (\ref{con}) we can differentiate
some different stages of the universe.
If
$u^2(t)={4 r^2 (b/b_0)^2 -1 \over 3} >1$, we obtain
that $r$ can take the values
$r > 1$ ($r < 1$) for
$b/b_0 < 1$ ($b/b_0 > 1$), respectively.
Note that now the condition (\ref{con}) implies that
$r\equiv r(t)$, for a given $h(t)$.
In this case $q < 0$, so that the expansion is accelerated.
On the other hand if
$u^2(t)={4 r^2 (b/b_0)^2 -1 \over 3} <1$, $r$ can take
the values $r< 1$ ($r > 1$) for
$b/b_0 > 1$ ($b/b_0 < 1$), respectively. In this
stage $q >0$ and the expansion of the universe is decelerated,
so that the function $u(t)$ take the values $0 < u(t) <1$
and the velocities (\ref{10}) become imaginary. Thus,
the metric (\ref{frw}) shifts its signature from $(+,-,-,-)$
to $(-,+,+,+)$.
When $u(t) =1$ the deceleration parameter
becomes zero because $\ddot b =0$.
At this moment the velocities (\ref{10})
rotates sincronically in the complex plane and
$r$ take values $r=1$ or $r<1$ for $b/b_0=1$ or $b/b_0 >1$, respectively.
On the other hand, the effective 4D
energy density operator $\rho$ is
\begin{equation}
\rho = \frac{1}{2}
\left[ \dot\varphi^2 + \frac{b^2_0}{b^2} \left(\nabla\varphi\right)^2
+2 V(\varphi)\right].
\end{equation}
Hence, the 4D expectation value of the Einstein equation
$\left(\dot b \over b\right)^2
= {8\pi G \over 3} \rho$ on the 4D FRW metric (\ref{frw}), will be
\begin{equation}
\left<h^2\right> = \frac{4\pi G}{3} \left< \dot\varphi^2 +
\frac{b^2_0}{b^2} \left(\nabla\varphi\right)^2 + 2 V(\varphi)\right>,
\end{equation}
where $G$ is the gravitational constant and $\left<H^2\right>
\equiv h^2=\dot b^2/b^2$.

In this frame the 5D momentum
$P^L$ is null: $P^L=0$.
This implies that the extra force will be
\begin{equation}
F_{ext} = 0.
\end{equation}
It also can be viewed from the point of view of the extra mass.
In this frame  $m^2_0 = M^2_{(5)}$ where
\begin{equation}
\left(\frac{\partial {\cal S}}{\partial\varphi_{,t}}\right)^2 -
e^{2\int h(t) dt}
\left(\frac{\partial {\cal S}}{\partial\varphi_{,R}}\right)^2=
M^2_{(5)}.
\end{equation}
Hence, the inertial 4D mass $m_0$ is the same than the invariant
5D mass $M_{(5)}$, so that there is not extra force on the
effective 4D frame. This can be justified from the fact that
the fifth coordinate $L$ do not varies in this frame. In other words
the 4D bulk $ds^2$ is the same that the 5D manifold $dS^2$, because
$dS_{(4)} =0$ for an observer that ``expands with the universe'' in
an inertial frame.

\subsection{A frame with variable fifth coordinate}\label{var}

Other interesting frame can be described by means of the transformation
$t = \int \psi(N) dN$, $R=r\psi $ and $\xi=\psi(N) e^{\int
\stackrel{\star}{H(N)}/H(N) dN}$, so that the 5D velocities
are
\begin{eqnarray}
&& U^t = \frac{2u(t)}{\sqrt{u^2(t) -1}}, \label{f1} \\
&& U^R = -\frac{2r}{\sqrt{u^2(t) -1}}, \label{f2} \\
&& U^{\xi} = \frac{u(t)}{\sqrt{u^2(t) -1}} \left(\frac{\dot H}{hH}
-\frac{\dot h}{h^2}\right) \frac{H}{H_0}. \label{f3}
\end{eqnarray}

In this frame
the 5D line element is given by
\begin{equation}\label{met}
dS^2 = \epsilon\left(
dt^2 - e^{2 \int h(t) dt} dR^2 - \left(\frac{H_0}{H}\right)^2 d\xi^2\right) ,
\end{equation}
where the 4D line element (or ``bulk'') $ds^2$ is given by
the first two terms in (\ref{met})
\begin{equation}\label{bulk}
ds^2 = \epsilon\left( dt^2 - e^{2 \int h(t) dt} dR^2\right),
\end{equation}
and $h^2(t) = H^2(t) + {C\over 3}$ for a given constant $C$.
Hence, the extra force on the 4D bulk
will be $F^{ext} = {dP^{\xi} \over ds}$.
Note that extra force becomes from the motion of the fifth coordinate
in the effective 4D bulk. In other words, an observer in the 4D bulk
(\ref{bulk}), will move under the influence of an extra force
that, in the example here studied, takes the form
\begin{equation}\label{for}
F^{ext} = 
\left(\left|1- \frac{r^2 \dot h^2}{h^4} e^{2\int h dt} \right|\right)^{-1/2}
\frac{dP^{\xi}}{dt},
\end{equation}
which is invariant under changes of signature (i.e., $\epsilon=1 \rightarrow
\epsilon=-1$).
The 5D Lagrangian in this frame takes the form
\begin{equation}\label{lag}
^{(5)}L(\varphi,\varphi_{,A})
= - \left(\frac{b}{b_0}\right)^3
\frac{H_0}{H}
\left(\frac{1}{2} g^{\alpha\beta} \varphi_{,\alpha}\varphi_{,\beta}
+ V(\varphi) \right),
\end{equation}
so that the momentum $P^{\xi}$ is
\begin{equation}
P^{\xi} = - \left(\frac{b}{b_0}\right)^3
\frac{H_0}{H} g^{\xi\xi}
\frac{\partial V(\varphi)}{\partial\varphi_{,\xi}}.
\end{equation}
In this representation the potential $V(\varphi)$ assumes the form
\begin{equation}\label{pote}
V(\varphi) = \frac{1}{2} \left[\left(R h\right)^2 -\left(\frac{b_0}{b}\right)^2
\right] \left(\nabla_{R} \varphi\right)^2 -
\frac{1}{2} \left(\frac{H}{H_0}\right)^2 \varphi^2_{,\xi} -
\frac{H}{H_0} \left(R h\right)\varphi_{,\xi} \nabla_{R} \varphi,
\end{equation}
so that the momentum $P^{\xi}$ is
\begin{equation}\label{p}
P^{\xi} =
\left(\frac{b}{b_0}\right)^3 \left[\left(\frac{H}{H_0} \right)
\varphi_{,\xi} + \left(R h\right) \nabla_{R}\varphi\right].
\end{equation}
Note that the effective kinetic component in the 5D Lagrangian
(\ref{lag}) is 4D, but the potential (\ref{pote}) is evaluated
in 5D frame (\ref{f1}),(\ref{f2}),(\ref{f3}).
From eqs. (\ref{for}) and (\ref{p}), we obtain the extra force for
this frame
\begin{eqnarray}
F^{ext}& = & \left(\frac{b}{b_0}\right)^3 \left[\left|1- \left(\frac{
R \dot h}{h}\right)^2 \left(\frac{b}{b_0}\right)^2\right| \right]^{-1/2}
\left[\left(3\frac{\dot b}{b} \frac{H}{H_0} + \frac{\dot H}{H} \right)
\varphi_{,\xi} + \left( 3 \frac{\dot b}{b} \left(R h\right) +
\left(\dot R h + R \dot h\right)\right) \nabla_{R}\varphi\right. \nonumber \\
&+& \left.\frac{H}{H_0} \frac{d}{dt}\left(\varphi_{,\xi}\right) +
\left(R h\right) \frac{d}{dt} \left(\nabla_r\varphi\right)\right], \label{fo}
\end{eqnarray}
where $\left({b\over b_0}\right)^2 = e^{2\int h dt}$.
The important fact here is that the extra force is originated in the
last two terms of the 5D potential (\ref{pote}), which depends on the fifth
coordinate $\xi$.

On the other hand the 4D squared mass of the inflation field $\varphi$
on the 4D bulk (\ref{bulk}), is given by
\begin{equation}
m^2_0 = \left(\frac{\partial {\cal S}}{\partial\varphi_{,t}}\right)^2 -
e^{-2\int h dt} \left(\frac{\partial {\cal S}}{\partial\varphi_{,R}}\right)^2,
\end{equation}
so that one obtains
\begin{equation}\label{ma}
m^2_0 - M^2_{(5)} = \left(\frac{H}{H_0}\right)^2
\left(\frac{\partial {\cal S}}{\partial\varphi_{,\xi}}\right)^2,
\end{equation}
which gives $m^2_0 \ge M^2_{(5)}$ because the right hand of the
equation (\ref{ma}) is positive (for $C >0$). This is an important
result which shows that the motion of the fifth coordinate has
an antigravitational effect on a observer in a 4D bulk in
which the inflaton field has a 4D mass $m_0$.
This fact should be responsible for the extra force (\ref{fo})
because the observer ``is placed'' in a non inertial
frame (or 4D bulk).
In this framework the motion of the fifth coordinate
is viewed on the bulk as an extra force.
Note that it becomes zero as $C \rightarrow 0$,
because in this limit $U^{\xi} \rightarrow
0$ and $V(\varphi) \rightarrow {1 \over 2} \left[
\left(R h\right)^2
-\left({b_0\over b}\right)^2 \right] \left(\nabla_{R} \varphi
\right)^2$.
On the other hand, $U^{\xi} \rightarrow 0$ as $t\rightarrow \infty$,
because $\dot H <0$ (and $\dot h <0$)
along all the history of the universe, such that
$\left({H\over H_0}\right)_{t\rightarrow \infty} \rightarrow 0$.
Hence, for very late times the external force (\ref{fo})
on the bulk becomes
negligible. However, this force should be very important in the early
universe when $H/H_0 \gg 1$ and
the equation of state is ${\rm p} \simeq -\rho$,
being ${\rm p}$ and $\rho$ respectively the pressure and energy density.
Note that $H_0$ is the value of the
Hubble parameter at the end of inflation.

To ilustrate the formalism we can consider the case where
$h(t) = t^{-1} \  p_1(t)$ and $H(t) = t^{-1} \  p(t)$, where
\begin{eqnarray}
&& p_1(t) = \sqrt{(2/3 + A t^{-2} - B t^{-1})^2 + \frac{C}{3} t^2}, \\
&& p(t) =  \sqrt{2/3 + A t^{-2} - B t^{-1}}
\end{eqnarray}
Here
$A=1.5 \  10^{30} \  {\rm G}^{1}$, $B=10^{15} \  {\rm G}^{1/2}$
and we take the special case where the constant
$C$ is the cosmological constant $\Lambda$: $\Lambda
= 1.5 \  10^{-121} \  {\rm G}^{-1}$.
Furthermore,
$G=M^{-2}_p$ is the gravitational constant and $M_p=1.2 \times
10^{19} \  {\rm GeV}$ is the
Planckian mass\cite{ultimo}.
The function $p_1(t)$ represents the power of expansion of the universe
in Planckian times. For early times $p_1(t) \gg 1$ and the
universe is accelerated ($\ddot b >0$).
At $t_r \simeq 10^{16} \  {\rm G^{1/2}}$,
after inflation ends, the universe becomes radiation dominated:
$p_1(t_r) \simeq 1/2$ and after it (for $t \gg t_r$) matter dominated:
$p_1(t \gg t_r) \simeq 2/3$. However, at $t \simeq 10^{60.22} \  {\rm G^{1/2}}$
the cosmological constant begins to be determinant and the universe
becomes almost vacuum dominated (quintessential expansion):
${\rm p} \lesssim -{2\over 3} \rho$. The function
$p(t)$ represents the power of expansion of the universe without the
cosmological constant. The difference between $p_1$
and $p$ becomes notorious for very large time (i.e., for $t
> 10^{60.22} \  {\rm G^{1/2}}$).
Numerical calculations give us the time for which $\ddot b =q=0$
at the end of inflation: $x(t_0)\simeq
14.778$ [we take
$x(t) = {\rm log}_{10}(t)$].
At this moment $N(t_0)=0$, but after it becomes positive.
Furthermore, for
$x(t) > x(t_*)$ [with $x(t_*) \simeq 60.22$], $p_1$
begins to increase from the value $p_1 \simeq 2/3$
and the 4D bulk universe is accelerated.
In other words, as we demonstrated in a prvious work\cite{ultimo}, the present
day observed (quintessential)
acceleration of the universe is consequence of the
nonzero $\Lambda$, and has been notorious since the universe was
nearly 4 billion years old (when the supernovae explosions ocurred).

\section{Final comments}

In this work we have
studied the possible origin of extra force and extra mass
from a noncompact Kaluza-Klein formalism
recently introduced by using the Hamilton-Jacobi formalism in the
framework of cosmological models.
However, it is one of the two possible outcomes to study this topic (see,
for example \cite{otra}).
We have examined the inertial
4D mass $m_0$ of the inflaton field
on a 4D FRW bulk in two examples.
In the first one $ds^2=dS^2$, so that the inertial mass $m_0$ is the same
than the 5D gravitational mass $M_{(5)}$ of the inflaton field.
As consequence of this fact there are no extra force on the 4D bulk.
However, in the second example
antigravitational effects on a non inertial 4D bulk
should be a consequence of the motion of the fifth coordinate with
respect to this bulk, because $dS^2 \neq ds^2$ so that $m^2_0 > M^2_{(5)}$.
This disagreement between the 4D inertial and 5D
gravitational masses is viewed
on the 4D bulk as an extra force. The important here is that $m_0$ has a
geometrical origin and depends on the frame of the observer. However,
$M_{(5)}$ is a 5D invariant gravitational mass and do not depends
on the frame of the observer.
In other words, all test particles travel on five-dimensional geodesics
but observers, who are bounded to spacetime, have access only to
the 4D part of the trajectory.
Finally, in the cosmological model here studied, we find that both,
the discrepance between $m_0$ and $M_{(5)}$ and extra force
(on cosmological scales), are bigger
in the early universe [i.e., during inflation
($x(t) < 14.778$)], but becomes negligible for large times.
However, from the point of view of experimentation, at present such
that discrepance should be more notorious on astrophysical scales, where
gravitational instabilities are important.\\

\vskip .2cm
\centerline{\bf{Acknowledgements}}
\vskip .2cm
The authors acknowledge Dr. J. Ponce de Leon for useful ideas.
JEMA acknowledges CONACyT and IFM of UMSNH (M\'exico)
for financial support.
MB acknowledges CONICET, AGENCIA 
and UNMdP (Argentina)
for financial support.\\

\end{document}